\documentclass[prl,aps,twocolumn,showpacs,superscriptaddress]{revtex4}

\usepackage{verbatim}
\usepackage{epsfig}
\usepackage{amssymb}
\usepackage{amsmath}
\usepackage{bm}

\begin{document}
\title{Evidence of Galaxy Cluster Motions with the Kinematic Sunyaev-Zel'dovich Effect}
\date{\today}

\author{Nick~Hand}
\affiliation{Department of Astronomy, University of California, Berkeley, CA 94720 USA} 
\affiliation{Department of Astrophysical Sciences, Princeton University, Princeton, NJ 08544 USA}
\author{Graeme~E.~Addison}
\affiliation{Sub-department of Astrophysics, University of Oxford, Denys Wilkinson Building, Keble Road, Oxford OX1 3RH UK}
\author{Eric~Aubourg}
\affiliation{APC, University of Paris Diderot, CNRS/IN2P3, CEA/IRFU, Observatoire de Paris, Sorbonne Paris Cit{\'e}, France}
\author{Nick~Battaglia}
\affiliation{Department of Physics, Carnegie Mellon University, Pittsburgh, PA 15213 USA}
\author{Elia~S.~Battistelli}
\affiliation{Department of Physics, University of Rome ``La Sapienza,'' Piazzale Aldo Moro 5, I-00185 Rome, Italy}
\author{Dmitry~Bizyaev}
\affiliation{Apache Point Observatory, P.O.\ Box 59, Sunspot, NM 88349 USA}
\author{J.~Richard~Bond}
\affiliation{Canadian Institute for Theoretical Astrophysics, University of Toronto, Toronto, ON M5S 3H8 Canada}
\author{Howard~Brewington}
\affiliation{Apache Point Observatory, P.O.\ Box 59, Sunspot, NM 88349 USA}
\author{Jon~Brinkmann}
\affiliation{Apache Point Observatory, P.O.\ Box 59, Sunspot, NM 88349 USA}
\author{Benjamin~R.~Brown}
\affiliation{Department of Physics and Astronomy, University of Pittsburgh, Pittsburgh, PA 15260 USA}
\author{Sudeep~Das}
\affiliation{Department of Physics, University of California, Berkeley CA 94720 USA}
\affiliation{Lawrence Berkeley National Laboratory, 1 Cyclotron Road, Berkeley, CA 94720 USA}
\author{Kyle~S.~Dawson}
\affiliation{Department of Physics and Astronomy, University of Utah, Salt Lake City, UT 84112 USA}
\author{Mark~J.~Devlin}
\affiliation{Department of Physics and Astronomy, University of Pennsylvania, Philadelphia, PA 19104 USA}
\author{Joanna~Dunkley}
\affiliation{Sub-department of Astrophysics, University of Oxford, Denys Wilkinson Building, Keble Road, Oxford OX1 3RH UK}
\author{Rolando~Dunner}
\affiliation{Departamento de Astronom{\'i}a y Astrof{\'i}sica, Pontificia Universidad Cat{\'o}lica de Chile, Santiago 22 Chile}
\author{Daniel~J.~Eisenstein}
\affiliation{Harvard College Observatory, 60 Garden Street, MS 20, Cambridge, MA 02138 USA}
\author{Joseph~W.~Fowler}
\affiliation{NIST Quantum Devices Group, 325 Broadway Mailcode 817.03, Boulder, CO 80305 USA}
\author{Megan~B.~Gralla}
\affiliation{Department of Physics and Astronomy, The Johns Hopkins University, Baltimore, MD 21218 USA}
\author{Amir~Hajian}
\affiliation{Canadian Institute for Theoretical Astrophysics, University of Toronto, Toronto, ON M5S 3H8 Canada}
\author{Mark~Halpern}
\affiliation{Department of Physics and Astronomy, University of British Columbia, Vancouver, BC V6T 1Z4 Canada}
\author{Matt~Hilton}
\affiliation{School of Physics and Astronomy, University of Nottingham, University Park, Nottingham, NG7 2RD UK}
\author{Adam~D.~Hincks}
\affiliation{Canadian Institute for Theoretical Astrophysics, University of Toronto, Toronto, ON M5S 3H8 Canada}
\affiliation{Joseph Henry Laboratories of Physics, Princeton University, Princeton, NJ 08544 USA}
\author{Ren{\'e}e~Hlozek}
\affiliation{Department of Astrophysical Sciences, Princeton University, Princeton, NJ 08544 USA}
\author{John~P.~Hughes}
\affiliation{Department of Physics and Astronomy, Rutgers University, Piscataway, NJ 08854 USA}
\author{Leopoldo~Infante}
\affiliation{Departamento de Astronom{\'i}a y Astrof{\'i}sica, Pontificia Universidad Cat{\'o}lica de Chile, Santiago 22 Chile}
\author{Kent~D.~Irwin}
\affiliation{NIST Quantum Devices Group, 325 Broadway Mailcode 817.03, Boulder, CO 80305 USA}
\author{Arthur~Kosowsky}
\affiliation{Department of Physics and Astronomy, University of Pittsburgh, Pittsburgh, PA 15260 USA}
\affiliation{Pittsburgh Particle Physics, Astrophysics, and Cosmology Center, University of Pittsburgh, Pittsburgh PA 15260}
\author{Yen-Ting~Lin}
\affiliation{Institute of Astronomy and Astrophysics, Academia Sinica, Taipei, Taiwan}
\author{Elena~Malanushenko}
\affiliation{Apache Point Observatory, P.O.\ Box 59, Sunspot, NM 88349 USA}
\author{Viktor~Malanushenko}
\affiliation{Apache Point Observatory, P.O.\ Box 59, Sunspot, NM 88349 USA}
\author{Tobias~A.~Marriage}
\affiliation{Department of Physics and Astronomy, The Johns Hopkins University, Baltimore, MD 21218 USA}
\author{Danica~Marsden}
\affiliation{Department of Physics, University of California, Santa Barbara, CA 93106 USA}
\author{Felipe~Menanteau}
\affiliation{Department of Physics and Astronomy, Rutgers University, Piscataway, NJ 08854 USA}
\author{Kavilan~Moodley}
\affiliation{Astrophysics and Cosmology Research Unit, School of Mathematical Sciences, 
University of KwaZulu-Natal, Durban 4041, South Africa}
\author{Michael~D.~Niemack}
\affiliation{NIST Quantum Devices Group, 325 Broadway Mailcode 817.03, Boulder, CO 80305 USA}
\author{Michael~R.~Nolta}
\affiliation{Canadian Institute for Theoretical Astrophysics, University of Toronto, Toronto, ON M5S 3H8 Canada}
\author{Daniel~Oravetz}
\affiliation{Apache Point Observatory, P.O.\ Box 59, Sunspot, NM 88349 USA}
\author{Lyman~A.~Page}
\affiliation{Joseph Henry Laboratories of Physics, Princeton University, Princeton, NJ 08544 USA}
\author{Nathalie Palanque-Delabrouille}
\affiliation{CEA, Centre de Saclay, IRFU, 91191 Gif-sur-Yvette, France}
\author{Kaike~Pan}
\affiliation{Apache Point Observatory, P.O.\ Box 59, Sunspot, NM 88349 USA}
\author{Erik~D.~Reese}
\affiliation{Department of Physics and Astronomy, University of Pennsylvania, Philadelphia, PA 19104 USA}
\author{David~J.~Schlegel}
\affiliation{Lawrence Berkeley National Laboratory, 1 Cyclotron Road, Berkeley, CA 94720 USA}
\author{Donald~P.~Schneider}
\affiliation{Department of Astronomy and Astrophysics, The Pennsylvania State University, University Park, PA 16802 USA}
\affiliation{Institute for Gravitation and the Cosmos, The Pennsylvania State University, University Park, PA 16802 USA}
\author{Neelima~Sehgal}
\affiliation{Department of Astrophysical Sciences, Princeton University, Princeton, NJ 08544 USA}
\author{Alaina~Shelden}
\affiliation{Apache Point Observatory, P.O.\ Box 59, Sunspot, NM 88349 USA}
\author{Jon~Sievers}
\affiliation{Joseph Henry Laboratories of Physics, Princeton University, Princeton, NJ 08544 USA}
\affiliation{Canadian Institute for Theoretical Astrophysics, University of Toronto, Toronto, ON M5S 3H8 Canada}
\author{Crist{\'o}bal~Sif{\'o}n}
\affiliation{Departamento de Astronom{\'i}a y Astrof{\'i}sica, Pontificia Universidad Cat{\'o}lica de Chile, Santiago 22 Chile}
\author{Audrey~Simmons}
\affiliation{Apache Point Observatory, P.O.\ Box 59, Sunspot, NM 88349 USA}
\author{Stephanie~Snedden}
\affiliation{Apache Point Observatory, P.O.\ Box 59, Sunspot, NM 88349 USA}
\author{David~N.~Spergel}
\affiliation{Department of Astrophysical Sciences, Princeton University, Princeton, NJ 08544 USA}
\author{Suzanne~T.~Staggs}
\affiliation{Joseph Henry Laboratories of Physics, Princeton University, Princeton, NJ 08544 USA}
\author{Daniel~S.~Swetz}
\affiliation{NIST Quantum Devices Group, 325 Broadway Mailcode 817.03, Boulder, CO 80305 USA}
\author{Eric~R.~Switzer}
\affiliation{Canadian Institute for Theoretical Astrophysics, University of Toronto, Toronto, ON M5S 3H8 Canada}
\author{Hy~Trac}
\affiliation{Department of Physics, Carnegie Mellon University, Pittsburgh, PA 15213 USA}
\author{Benjamin~A.~Weaver}
\affiliation{Center for Cosmology and Particle Physics, New York University, New York, NY 10003 USA}
\author{Edward~J.~Wollack}
\affiliation{Code 553/665, NASA/Goddard Space Flight Center, Greenbelt, MD 20771 USA}
\author{Christophe Yeche}
\affiliation{CEA, Centre de Saclay, IRFU, 91191 Gif-sur-Yvette, France}
\author{Caroline~Zunckel}
\affiliation{Astrophysics and Cosmology Research Unit, School of Mathematical Sciences, 
University of KwaZulu-Natal, Durban 4041, South Africa}

\begin{abstract}
Using high-resolution microwave sky maps made by the Atacama Cosmology Telescope, we
for the first time present strong evidence for motions of galaxy clusters and groups via microwave background
temperature distortions due to the kinematic Sunyaev-Zel'dovich effect.
Galaxy clusters are identified by their constituent luminous galaxies observed by the 
Baryon Oscillation Spectroscopic Survey, part of the Sloan
Digital Sky Survey III. We measure the mean pairwise momentum of clusters,
with a probability of the signal being due to random errors of 0.002,
and the signal is consistent with the growth of cosmic
structure in the standard model of cosmology. 
\end{abstract}


\maketitle

{\it Introduction.} The growth of cosmic structure over the history of the Universe inevitably results not only
in the formation of dense objects, but also in motions of these objects. Measurements of these motions
have the potential to provide both a valuable consistency check on the standard cosmological model, and also
an independent route to constraining cosmological parameters and the nature of dark energy. 

In  1972, Sunyaev and Zel'dovich realized that a moving galaxy cluster, which is largely composed
of hot, ionized gas in a dark matter potential well, will induce a small brightness temperature shift in the
microwave radiation passing through it. The shift is proportional to both the mass in electrons and
the line-of-sight velocity of the cluster with respect to the microwave background rest frame 
\cite{sun72,sun80}. This kinematic Sunyaev-Zel'dovich (kSZ) effect is distinct from
the thermal SZ  (tSZ) effect, in which scattering from the same hot cluster gas creates
a spectral distortion (see \cite{car02} for a review). In high-mass clusters ($M\simeq 10^{15}$ solar masses),
the tSZ signal is typically a factor of 20 larger than the kSZ signal; however, the two signals are comparable
for the low-mass clusters ($M\simeq 10^{13}\, M_\odot$) which are far more abundant. 
(For brevity, we refer to any object with mass larger than $10^{13}\, M_\odot$  as a cluster, even though
objects below $10^{14}\, M_\odot$ are usually referred to as ``groups''.)
The tSZ effect from large clusters is now regularly 
observed in blind surveys \cite{mar11b,wil11,hin10,van10}, 
but only upper limits for the kSZ effect from
individual galaxy clusters have been achieved to date \cite{hol97,mau00,ben03,zem03,mau12}. 

In this paper, we present clear statistical evidence of the motions of galaxy clusters through their
kSZ signal in arcminute-resolution microwave maps made with the Atacama Cosmology Telescope (ACT) \cite{swe11}. 
Luminous galaxies 
are associated with galaxy clusters \cite{eis01,ho09}, and we use the Sloan Digital Sky Survey III (SDSS-III) Baryon Oscillation Spectroscopic Survey (BOSS) \cite{eis11}
catalog of these galaxies as galaxy
cluster proxies, giving the sky location and redshift for thousands of potential clusters. We then treat the effective microwave
temperature at 148 GHz measured by ACT in the direction of  the cluster as a noisy estimator of the cluster's line-of-sight momentum, due
to the kSZ temperature shift from that cluster.
Individual cluster momentum measurements have a low signal-to-noise ratio, but we combine a large number of
differential measurements to obtain estimates of the mean relative momentum of cluster pairs in bins of
comoving cluster separation. This statistic is insensitive to the tSZ signal, galaxy emission, and other sky signals in
the ACT data. 
The conventional scenario of structure formation, driven by gravitational attraction,
predicts that any pair of clusters should have a slight tendency to be moving towards each other
rather than away from each other \cite{peebles,dia00}, and we see the expected signal in our data at a 
high statistical significance.

{\it Survey Data Sets.} We make use of two astronomical survey data sets. The first is a 148 GHz sky map from
ACT, a dedicated microwave survey telescope in the Atacama Desert
of Chile. The map 
covers a strip approximately $3^\circ$ wide and $110^\circ$ long with an angular resolution 
of $1.4^\prime$, centered on the celestial equator and obtained
over three observing seasons from 2008 to 2010 \cite{swe11}. 
We match-filter the ACT 148 GHz map with a characteristic filter scale at the 
map resolution of $1.4^\prime$ \cite{han11}, to suppress noise from
the primary microwave background fluctuations.
Filtered map pixels are $0.5^\prime$ square, and have a noise per pixel ranging from 15 to 25 $\mu$K brightness temperature;  
the map is calibrated to 2\% by comparing with WMAP \cite{haj11}. Imperfect knowledge of the beam profile gives an additional 1\% calibration uncertainty on galaxy cluster scales.
(A similar map at 218 GHz has higher noise and is used only for Table 1 below.) 

\begin{table*}[t]
\begin{center}
\begin{tabular}{| c | c | c | c | c |  c | c | c |}
\hline
Bin & \phantom{n}$N_{\mathrm{gal}}$\phantom{n}  &  $ \langle L_{0.1r} \rangle$  & $L_{0.1r}$ Range   & 
\phantom{m}$\langle z \rangle $\phantom{m} & $\delta T_{\mathrm{148}}$ & $\delta T_{\mathrm{218}}$  & $\delta T_{\mathrm{tSZ}}$  \\
       &                  & $10^{10}L_{\odot}$  &  $10^{10}L_{\odot}$  &   & $\mu$K  & $\mu$K   & $\mu$K \\  \hline
1 & \phantom{14}225    & 23.3 &  17.7\phantom{1} -- 73.5  &  0.65  & $-6.98 \pm 1.69$  & $+1.35 \pm 2.59$  & $-7.42 \pm 1.89$ \\
2 & \phantom{1}1326    & 13.1  & 11.0\phantom{1} -- 17.7   &  0.61  & $-1.33 \pm 0.72$  & $+3.46 \pm 1.06$  & $-2.45 \pm 0.80$  \\
3 & \phantom{1}4100  &  \phantom{1}9.0   &  \phantom{1}7.8\phantom{1} -- 11.0 & 0.57  & $-0.11 \pm 0.38$ & $+2.16 \pm 0.60$   & $-0.81 \pm 0.43$\\
4 & \phantom{1}8467 & \phantom{1}6.6  &  \phantom{1}5.7\phantom{1} -- \phantom{1}7.8  &  0.51    & $+0.35 \pm 0.28$ & $+2.17 \pm 0.41$  & $-0.36 \pm 0.31$ \\
5 & 13173 & \phantom{1}4.3 & \phantom{1}0.01 -- \phantom{1}5.7 & 0.48 & $+0.43 \pm 0.22 $ & $+1.53 \pm 0.33$  & $-0.07 \pm 0.24$ \\ \hline \hline 
total & 27291 & 6.3 & \phantom{1}0.01 -- 73.5 & 0.51 & $+0.17 \pm 0.15 $ & $+1.92 \pm 0.22$  & $-0.45 \pm 0.17$  \\
\hline
\end{tabular}
\end{center}
\caption{Mean brightness temperature fluctuations in sky directions corresponding to the BOSS DR9 galaxies, in
bins of galaxy luminosity. The right
column corresponds to the tSZ brightness temperature at 148 GHz, $\delta T_{\rm tSZ}\equiv \delta T_{148} - 0.325\, \delta T_{218}$,
projecting out a dust emission component \cite{hal10}.
The ACT maps
are match-filtered at an angular scale of $1.4^\prime$, equal to the beam size at 148 GHz, then
subpixelized, convolved with the beam profile, and summed over all subpixels within $4^{\prime\prime}$ of the galaxy. }
\end{table*}

The second data set is a catalog of luminous galaxies from BOSS Data Release 9 (a combination of the 
CMASS and LOWZ samples from DR9), 
a component of the Sloan Digital Sky Survey III \cite{fuk96,gun98,gun06}. The catalog contains 27291 galaxies in
a 220-square-degree region overlapping the ACT sky region
(right ascension range $-43^\circ$ to $+45^\circ$). Galaxies are selected to lie at least $1^\prime$ away from 
any radio source in the 1.4 GHz FIRST radio catalog \cite{FIRST}; radio contamination is not a significant issue. 
Spectroscopic redshifts range from $z=0.05$ to $z=0.8$ with a mean redshift of 0.51. Luminosities are estimated 
as in Ref.~\cite{han11}, with an additional mean luminosity evolution correction as given in Ref.~\cite{teg04}.
A halo-model correlation function analysis shows that most of the BOSS galaxies
reside in haloes with masses around $10^{13}$ solar masses, with around 10\% to 15\% in haloes as large
as $10^{14}\, M_\odot$ \cite{whi11}. 

To estimate the microwave temperature distortion $T_i$ associated with galaxy $i$,
we follow the procedure used in
Ref.~\cite{han11}: a $10^\prime$ by $10^\prime$ submap centered on the galaxy is repixellized into
$0.0625^\prime$ subpixels, convolved with the ACT beam profile to smooth the map, and then
averaged over all subpixels within $1^\prime$ of the galaxy. The $1^\prime$ binning radius 
maximizes the signal-noise ratio of our kSZ detection, but varying the binning radius
between $4^{\prime\prime}$ and $4^\prime$ only marginally changes the detection significance. 

The most luminous galaxies in our catalog have the largest halo masses. 
To confirm this, we divide our sample into five luminosity bins; 
Table 1 displays the mean central temperature distortion corresponding to the galaxies in each bin.
The rightmost column gives the tSZ distortion 
brightness temperature at 148 GHz, with the next-largest component due to dust emission
projected out \cite{hal10}; this is obtained from a linear combination of ACT 148 and
218 GHz signals. Galaxies in the three highest luminosity bins, corresponding to about 20\% of the total, show 
mean temperature decrements consistent with halo-model cluster masses \cite{whi11}
and with the mean temperature decrements found in Ref.~\cite{han11}. 

{\it Mean Pairwise Momentum.} 
Combining the above survey data provides a set of galaxy cluster sky positions and redshifts (from the luminous 
galaxy positions and redshifts) and line-of-sight momenta (from the ACT temperature). To compare this data set
with cosmological models, consider the mean pairwise momentum statistic:
\begin{equation}
p_{\rm pair}(r) \equiv \left\langle( {\bf p}_i - {\bf p}_j)\cdot{\bf\hat r}_{ij}\right\rangle,
\label{ppair_def}
\end{equation}
where galaxy cluster $i$ has momentum ${\bf p}_i$ and comoving position ${\bf r}_i$, 
the comoving separation vector between a pair of
clusters $i$ and $j$ is ${\bf r}_{ij}\equiv {\bf r}_i - {\bf r}_j$, overhats denote unit vectors, and the average on the right side
of the equation is over all cluster pairs in a bin around comoving separation $r\equiv \left|{\bf r}_{ij}\right|$. If two galaxy
clusters are moving towards each other, their contribution to $p_{\rm pair}(r)$ will be negative, and if moving
apart, positive. 
An estimator of $p_{\rm pair}(r)$ using only line-of-sight momenta is \cite{fer99}
\begin{eqnarray}
&&{\tilde p}_{\rm pair}(r) = \frac{\sum_{i<j}({\bf p}_i \cdot{\bf\hat r}_i - {\bf p}_j \cdot{\bf\hat r}_j)c_{ij}}
{\sum_{i<j} c_{ij}^2}\label{ppair_est}\\
&&c_{ij} \equiv {{\bf\hat r}_{ij}}\cdot
\frac{{\bf\hat r}_i + {\bf\hat r}_j}{2}
= \frac{(r_i - r_j)(1+\cos\theta)}{2\sqrt{r_i^2 + r_j^2 - 2r_ir_j\cos\theta}},\qquad
\label{cij_def}
\end{eqnarray}
where $\theta$ is the angular separation between two clusters on the sky and $r_i\equiv\left|{\bf r}_i\right|$ 
is the comoving distance to cluster $i$, which can be computed 
from the cluster redshift using standard $\Lambda$CDM cosmological parameters \cite{dun11}.  
(The cluster velocity gives 
negligible contribution to the distance estimate for clusters at a cosmological distance, and a bias in estimating
Eq.~(\ref{ppair_est}) which is small compared to the measurement errors.)
The statistic $p_{\rm pair}(r)$ is equal to the familiar mean pairwise velocity $v_{\rm pair}(r)$ \cite{dav77,jus00,fel03}
times the average mass of the clusters in the sample.

We can measure the line-of-sight component of the
momentum via the kSZ microwave temperature fluctuation,  $ T_{{\rm kSZ},i} \equiv -N_{\rm kSZ}\, {\bf p}_i\cdot{\bf\hat r}_i$,
assuming that the ratio of the total cluster mass to its mass
in hot gas is simply the universal ratio of matter density $\Omega_m$ to baryon density $\Omega_b$ \cite{bha08}.
The normalization $N_{\rm kSZ}$ depends on the pixel scale and beam size of the microwave map, and the cluster
density profile. Simulations including these effects \cite{seh10} give an expected mean temperature signal in the ACT 148 GHz map
of 1.6 and 0.3 $\mu$K  for clusters with a typical line-of-sight velocity of 200 km/s and masses $10^{14}$ and $10^{13}$ $M_\odot$.

The statistic $p_{\rm pair}(r)$ is both linear and differential, giving it desirable systematic error properties \cite{bk08b}. Any microwave temperature
signal associated with individual galaxy clusters, like the tSZ effect, 
will average to zero as long as it does not depend on the
relative distance between cluster pairs. Redshift-dependent signals can contribute to $p_{\rm pair}(r)$ and be confused with
the cluster kSZ signal, including infrared emission from galaxies in the cluster (which increases
with redshift out to $z=2$), any radio source emission, and small variations of the tSZ signal due
to evolution of average cluster mass and temperature. However, we can measure these effects on average
by simply finding the average microwave temperature ${\cal T}(z)$ corresponding to clusters at a given redshift,
and correcting the temperature in the direction of an individual galaxy cluster for this redshift-dependent piece. We evaluate 
a smoothed ${\cal T}(z)$ by averaging the
temperature towards all galaxies, each with redshift $z_i$ and a Gaussian weight factor $\exp[-(z-z_i)^2/2\sigma_z^2]$
with $\sigma_z=0.01$; our results are nearly insensitive to the value of $\sigma_z$ within a wide range.  
The resulting ${\cal T}(z)$ has a mean near zero and  an absolute value of up to $3\,\mu$K. 

\begin{figure}[t]
\begin{center}
\includegraphics[width=3.4in]{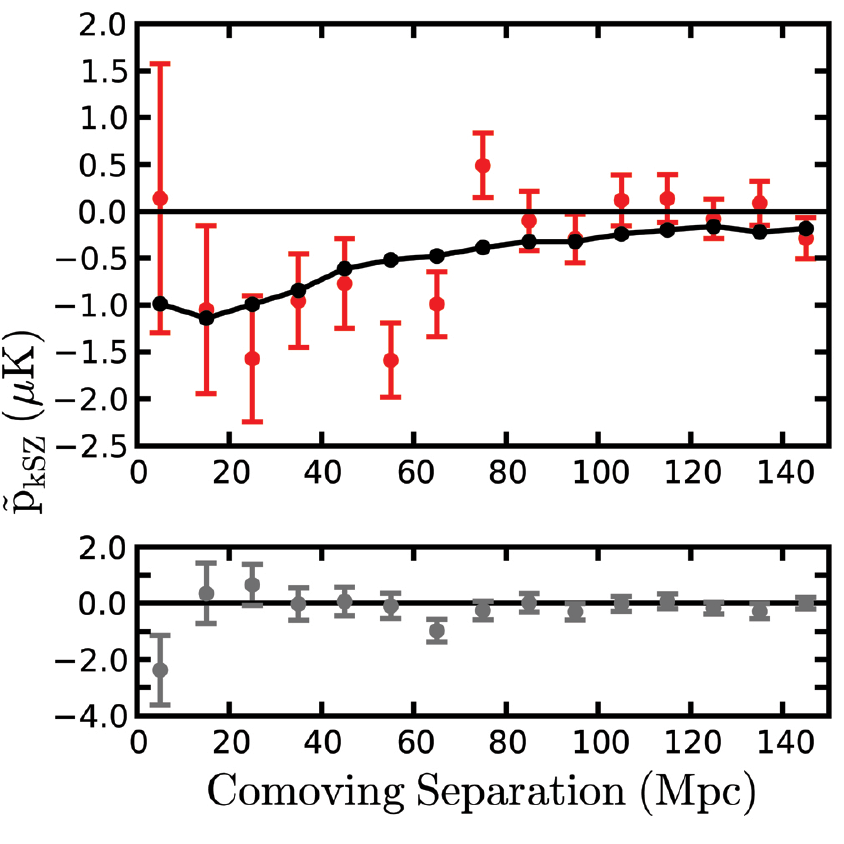}
\end{center}
\caption{The upper panel shows the mean pairwise momentum estimator, Eq.~(\ref{ppair_eval}), for the 5000 most luminous
BOSS DR9 galaxies within the ACT sky region (red points), with bootstrap errors. 
The solid line is derived from numerical kSZ simulations \cite{seh10} using a halo
mass cutoff of $M_{200}=4.1\times 10^{13}\,M_\odot$.
The probability of the data given a null signal is $2.0\times 10^{-3}$ including bin covariances. 
The lower panel displays the same sum but with randomized map positions, and is consistent with a null signal. 
}
\label{fig:ppair}
\end{figure}

We thus evaluate the mean pairwise kSZ signal, correcting for possible redshift-dependent temperature
contributions, as
\begin{equation}
{\tilde p}_{\rm kSZ}(r) = -\frac{\sum_{i<j}\left[(T_i - {\cal T}(z_i)) - (T_j - {\cal T}(z_j))\right]c_{ij}}
{\sum_{i<j} c_{ij}^2}.
\label{ppair_eval}
\end{equation}
This quantity differs from  Eq.~(\ref{ppair_est}) by the amplitude factor $N_{\rm kSZ}$.
Figure \ref{fig:ppair} displays this statistic for the ACT pixel temperatures corresponding to the
5000 most luminous BOSS DR9 galaxies in the ACT sky region ($L > 8.1\times10^{10}\,L_\odot$); 
this luminosity cut minimizes the total noise from combined Poisson and pixel noise. 
Also displayed is
the signal extracted from a kSZ-only sky simulation, based on underlying large-volume
cosmological simulations \cite{seh10}, adjusting the mass limit of the simulation halos
to give the best fit to the data. We infer that our galaxy luminosity cut corresponds
to a cluster halo mass limit of roughly $M_{200}\simeq 4.1\times 10^{13}\,M_\odot$ and a mean
cluster halo mass of $M_{200}=6.5\times 10^{13}\,M_\odot$.
Error bars are estimated via bootstrap resampling. 
Neighboring bins have a mean correlation of 0.25 and we include smaller mean correlations out
to a 5-bin separation, as determined using independent simulation volumes.
 
The measured points largely fall below zero and have $\Delta\chi^2=23$ for 15 degrees of freedom,
compared to the best-fit model. The model is a good fit to the data:
13\% of random data realizations with the same normal errors and correlations 
have larger $\Delta\chi^2$. The measured points have $\Delta\chi^2=43$ for 15 degrees of freedom,
compared to a null signal; the probability of random noise having $\Delta\chi^2$ at least this large
is $2.0\times 10^{-3}$ including correlations. The measured points approach
zero signal as the comoving pair separation increases, which demonstrates that the signal depends on
spatial separation, not redshift separation.

Null tests are simple, as the statistic is essentially a sum of pixel temperatures, half with positive and half with
negative signs, with weights corresponding to relative galaxy positions.  Figure~\ref{fig:ppair} also displays 
the null test corresponding to using the same weights but random pair positions compared to the signal plot
($\Delta\chi^2=11.6$ for 15 degrees of freedom).
Success of this null test
verifies that the function ${\cal T}(z)$ correctly models any redshift-dependent contributions to the microwave signal. 
Changing the sign in the second term of Eq.~(\ref{ppair_eval}) from negative to positive also gives a null signal
($\Delta\chi^2=9.9$ for 15 degrees of freedom).

{\it Discussion and Prospects.}
The signal in Fig.~\ref{fig:ppair} represents the first measurement of the cosmic
velocity field made directly with respect to the rest frame of the Universe. It is consistent
with simulations based on the standard cosmological model. This signal is also the first
clear evidence for the kinematic Sunyaev-Zel'dovich effect. A recent 
attempt by Kashlinsky et al.\ to measure the
large-scale bulk flow via the galaxy cluster kSZ signal uses galaxy clusters
from X-ray surveys and searches for an overall dipole dependence of the microwave temperature
in the WMAP data at these locations \cite{kas08,kas11}. However, Keisler \cite{kei09} found the
first reported detection was not statistically significant. Osborne et al.~\cite{osb11} reanalyzed the most recent
results including both a monopole and dipole term, obtaining limits on a bulk flow a factor of
three below the reported detection of Ref.~\cite{kas11}. Mody and Hajian \cite{mod12} also fail to reproduce the
bulk flow result using Planck and ROSAT galaxy clusters. Planck will soon make a more precise test of this
reported large-scale flow \cite{mak11}. The statistic used in this paper is differential, which mitigates many of the
potential systematic errors affecting bulk flow measurements, but also is not sensitive to an overall bulk flow. 

Most previous work on peculiar velocities using optical observations has measured
the properties of the local bulk flow, but has not been able to extend measurements to cosmologically
interesting distances. The traditional method of measuring velocities 
-- a Doppler shift of an object's radiation spectrum -- is very challenging at cosmological distances because the spectrum of an object is redshifted due to the expansion of the Universe, and this cosmological redshift is typically large compared to the velocity frequency shift. Precise distance measurements are required, a difficult observational problem. Recent optical work
\cite{fel10} extends to around 100 Mpc, a redshift of $z=0.02$, while this paper uses galaxy cluster velocities out
to $z=0.8$. Future large optical surveys such as the LSST may enable competitive cosmological
velocity surveys using large catalogs of standard candles for distance measurements \cite{bha11}.

The evidence for a nonzero mean pairwise momentum from a kSZ signal presented here 
can also be interpreted as a measure of baryons on cluster length scales; a deficit of observed
baryons has long been a cosmological puzzle \cite{bre07}.  Our signal is roughly consistent with
the standard baryon fraction based on primordial nucleosynthesis, given
independent halo mass estimates based on clustering of our luminous galaxy sample. This issue will be addressed in
a future paper. 

Future improved measurements of the mean pairwise velocity have the potential to put strong
constraints on dark energy and modified gravity \cite{bk07,bk08a,bk09}. 
The measurement we have presented here is the
first step on a new path to constraining structure growth in the Universe.

\begin{acknowledgments}
This work was supported by the U.S.\ National Science Foundation through Grants AST-0408698 for the ACT project, and PHY-0355328, AST-0707731 and PIRE-0507768 (Grant number OISE-0530095). The PIRE program made possible exchanges between Chile, South Africa, Spain and the US that enabled this research program. NH was partly supported by the
Berkeley Fellowship for Graduate Study. AK was partly supported
by NSF grant AST-0807790. Funding was also provided by Princeton University and the University of Pennsylvania. ACT mapmaking computation was performed on the GPC supercomputer at the SciNet HPC Consortium; SciNet is funded by the Canada Foundation for Innovation under the auspices of Compute Canada, the Government of Ontario, Ontario
Research Fund -- Research Excellence, and the University of Toronto.
ACT operates in the Chajnantor Science Preserve in northern Chile under the
auspices of the Comisi{\'o}n Nacional de Investigaci{\'o}n Cientifica y Tecnol{\'o}gica (CONICYT).
This work made use of the NASA Astrophysical Data System for bibliographic information. 

Funding for SDSS-III has been provided by the Alfred P. Sloan Foundation, the Participating Institutions, the National Science Foundation, and the U.S. Department of Energy Office of Science. The SDSS-III web site is http://www.sdss3.org/.
SDSS-III is managed by the Astrophysical Research Consortium for the Participating Institutions of the SDSS-III Collaboration including the University of Arizona, the Brazilian Participation Group, Brookhaven National Laboratory, University of Cambridge, Carnegie Mellon University, University of Florida, the French Participation Group, the German Participation Group, Harvard University, the Instituto de Astrofisica de Canarias, the Michigan State/Notre Dame/JINA Participation Group, Johns Hopkins University, Lawrence Berkeley National Laboratory, Max Planck Institute for Astrophysics, New Mexico State University, New York University, Ohio State University, Pennsylvania State University, University of Portsmouth, Princeton University, the Spanish Participation Group, University of Tokyo, University of Utah, Vanderbilt University, University of Virginia, University of Washington, and Yale University.

We dedicate this paper to the memory of Roman Juszkiewicz (1952--2012), a pioneer in mean pairwise velocities and
a kSZ effect enthusiast.

\end{acknowledgments}


\end{document}